\documentclass[twocolumn,showpacs,preprintnumbers,amsmath,amssymb,superscriptaddress]{revtex4-1}

\usepackage{graphicx}
\usepackage{dcolumn}
\usepackage{bm}

\begin{document}

\title{Improved measurement of the longitudinal spin transfer to $\Lambda$ and $\bar \Lambda$ hyperons \\ in polarized proton--proton collisions at $\sqrt s$ = 200\,GeV}

\affiliation{Abilene Christian University, Abilene, Texas   79699}
\affiliation{AGH University of Science and Technology, FPACS, Cracow 30-059, Poland}
\affiliation{Alikhanov Institute for Theoretical and Experimental Physics, Moscow 117218, Russia}
\affiliation{Argonne National Laboratory, Argonne, Illinois 60439}
\affiliation{Brookhaven National Laboratory, Upton, New York 11973}
\affiliation{University of California, Berkeley, California 94720}
\affiliation{University of California, Davis, California 95616}
\affiliation{University of California, Los Angeles, California 90095}
\affiliation{University of California, Riverside, California 92521}
\affiliation{Central China Normal University, Wuhan, Hubei 430079 }
\affiliation{University of Illinois at Chicago, Chicago, Illinois 60607}
\affiliation{Creighton University, Omaha, Nebraska 68178}
\affiliation{Czech Technical University in Prague, FNSPE, Prague 115 19, Czech Republic}
\affiliation{Technische Universit\"at Darmstadt, Darmstadt 64289, Germany}
\affiliation{E\"otv\"os Lor\'and University, Budapest, Hungary H-1117}
\affiliation{Frankfurt Institute for Advanced Studies FIAS, Frankfurt 60438, Germany}
\affiliation{Fudan University, Shanghai, 200433 }
\affiliation{University of Heidelberg, Heidelberg 69120, Germany }
\affiliation{University of Houston, Houston, Texas 77204}
\affiliation{Indiana University, Bloomington, Indiana 47408}
\affiliation{Institute of Modern Physics, Chinese Academy of Sciences, Lanzhou, Gansu 730000 }
\affiliation{Institute of Physics, Bhubaneswar 751005, India}
\affiliation{University of Jammu, Jammu 180001, India}
\affiliation{Joint Institute for Nuclear Research, Dubna 141 980, Russia}
\affiliation{Kent State University, Kent, Ohio 44242}
\affiliation{University of Kentucky, Lexington, Kentucky 40506-0055}
\affiliation{Lawrence Berkeley National Laboratory, Berkeley, California 94720}
\affiliation{Lehigh University, Bethlehem, Pennsylvania 18015}
\affiliation{Max-Planck-Institut f\"ur Physik, Munich 80805, Germany}
\affiliation{Michigan State University, East Lansing, Michigan 48824}
\affiliation{National Research Nuclear University MEPhI, Moscow 115409, Russia}
\affiliation{National Institute of Science Education and Research, HBNI, Jatni 752050, India}
\affiliation{National Cheng Kung University, Tainan 70101 }
\affiliation{Nuclear Physics Institute AS CR, Prague 250 68, Czech Republic}
\affiliation{Ohio State University, Columbus, Ohio 43210}
\affiliation{Institute of Nuclear Physics PAN, Cracow 31-342, Poland}
\affiliation{Panjab University, Chandigarh 160014, India}
\affiliation{Pennsylvania State University, University Park, Pennsylvania 16802}
\affiliation{Institute of High Energy Physics, Protvino 142281, Russia}
\affiliation{Purdue University, West Lafayette, Indiana 47907}
\affiliation{Pusan National University, Pusan 46241, Korea}
\affiliation{Rice University, Houston, Texas 77251}
\affiliation{Rutgers University, Piscataway, New Jersey 08854}
\affiliation{Universidade de S\~ao Paulo, S\~ao Paulo, Brazil 05314-970}
\affiliation{University of Science and Technology of China, Hefei, Anhui 230026}
\affiliation{Shandong University, Qingdao, Shandong 266237}
\affiliation{Shanghai Institute of Applied Physics, Chinese Academy of Sciences, Shanghai 201800}
\affiliation{Southern Connecticut State University, New Haven, Connecticut 06515}
\affiliation{State University of New York, Stony Brook, New York 11794}
\affiliation{Temple University, Philadelphia, Pennsylvania 19122}
\affiliation{Texas A\&M University, College Station, Texas 77843}
\affiliation{University of Texas, Austin, Texas 78712}
\affiliation{Tsinghua University, Beijing 100084}
\affiliation{University of Tsukuba, Tsukuba, Ibaraki 305-8571, Japan}
\affiliation{United States Naval Academy, Annapolis, Maryland 21402}
\affiliation{Valparaiso University, Valparaiso, Indiana 46383}
\affiliation{Variable Energy Cyclotron Centre, Kolkata 700064, India}
\affiliation{Warsaw University of Technology, Warsaw 00-661, Poland}
\affiliation{Wayne State University, Detroit, Michigan 48201}
\affiliation{Yale University, New Haven, Connecticut 06520}

\author{J.~Adam}\affiliation{Creighton University, Omaha, Nebraska 68178}
\author{L.~Adamczyk}\affiliation{AGH University of Science and Technology, FPACS, Cracow 30-059, Poland}
\author{J.~R.~Adams}\affiliation{Ohio State University, Columbus, Ohio 43210}
\author{J.~K.~Adkins}\affiliation{University of Kentucky, Lexington, Kentucky 40506-0055}
\author{G.~Agakishiev}\affiliation{Joint Institute for Nuclear Research, Dubna 141 980, Russia}
\author{M.~M.~Aggarwal}\affiliation{Panjab University, Chandigarh 160014, India}
\author{Z.~Ahammed}\affiliation{Variable Energy Cyclotron Centre, Kolkata 700064, India}
\author{I.~Alekseev}\affiliation{Alikhanov Institute for Theoretical and Experimental Physics, Moscow 117218, Russia}\affiliation{National Research Nuclear University MEPhI, Moscow 115409, Russia}
\author{D.~M.~Anderson}\affiliation{Texas A\&M University, College Station, Texas 77843}
\author{R.~Aoyama}\affiliation{University of Tsukuba, Tsukuba, Ibaraki 305-8571, Japan}
% ...

\collaboration{STAR Collaboration}\noaffiliation

\date{\today}

\begin{abstract}
The longitudinal spin transfer $D_{LL}$ to $\Lambda$ and $\bar{\Lambda}$ hyperons produced in high-energy polarized proton--proton collisions is expected to be sensitive to the helicity distribution functions of strange quarks and anti-quarks of the proton, and to longitudinally polarized fragmentation functions. We report an improved measurement of $D_{LL}$ from data obtained at a center-of-mass energy of $\sqrt{s}$ = 200\,GeV with the STAR detector at RHIC. The data have an approximately twelve times larger figure-of-merit than prior results and cover $|\eta|<$ 1.2 in pseudo-rapidity with transverse momenta $p_T$ up to 6\,GeV/$c$.
In the forward scattering hemisphere at largest $p_T$, the longitudinal spin transfer is found to be $D_{LL}$ = -0.036 $\pm$ 0.048 (stat) $\pm$ 0.013(sys) for $\Lambda$ hyperons and $D_{LL}$ = 0.032 $\pm$ 0.043\,(stat) $\pm$ 0.013\,(sys) for $\bar{\Lambda}$ anti-hyperons.
The dependences on $\eta$ and $p_T$ are presented and compared with model evaluations.
\end{abstract}
\maketitle

The self-analyzing weak decay of $\Lambda$, $\bar{\Lambda}$, and other hyperons makes it possible to study a number of spin phenomena in nature.  In high-energy collisions of heavy ion beams, for example, a substantial alignment was recently observed between the angular momentum of the colliding system and the spin of the emitted hyperons~\cite{STAR:2017ckg}.
This provides a new way to study the hot and dense matter produced in such collisions.  The discovery of substantial transverse polarization in inclusive $\Lambda$ production at forward rapidities by protons on nuclear targets continues to present a challenge for theoretical models~\cite{Bunce:1976yb}.  Sizable longitudinal $\Lambda + \bar{\Lambda}$ polarization effects have been observed in $e^+ +\,e^-$ annihilation at an energy corresponding to the $Z^0$ pole~\cite{Buskulic:1996vb,Ackerstaff:1997nh}, originating mostly from fragmentation of the strongly polarized strange quarks and anti-quarks from $Z^0$ decay~\cite{Boros:1998kc,deFlorian:1997zj,Liu:2000fi}.  The spin transfer to the struck quarks is expected to play an important role in semi-inclusive deep-inelastic scattering spin-transfer measurements of longitudinally polarized positron and muon beams off unpolarized targets~\cite{Adams:1999px, Airapetian:2006ee,Alekseev:2009ab}, while 
neutrino measurements~\cite{Astier:2000ax,Astier:2001ve} are sensitive to fragments of the target remnant.

In longitudinally polarized $p + p$ collisions, the spin transfer $D_{LL}$ to a $\Lambda$ hyperon is defined as:
\begin{equation}
D_{LL}\equiv \frac
{\sigma_{p^+p \to  \Lambda ^+ X}-\sigma_{p^+p \to  \Lambda ^-X}}
{\sigma_{p^+p \to  \Lambda ^+ X}+\sigma_{p^+p \to  \Lambda ^-X}},
\label{gener1}
\end{equation}
where $\sigma$ denotes the (differential) production cross-section and the superscripts $+$ or $-$ denote the helicity of the beam proton or the produced $\Lambda$ hyperon.
The spin transfer for $\bar{\Lambda}$ is defined similarly.
At hard scales, $D_{LL}$ is sensitive to the internal spin structure of the proton and of the $\Lambda$ or $\bar{\Lambda}$ hyperon.
Theory expectations~\cite{deFlorian:1998ba,Boros:2000ya,Ma:2001na,Xu:2002hz,Xu:2005ru} describe $D_{LL}$ in factorized frameworks, where it then arises from quark and anti-quark parton distribution functions (PDFs) in the polarized proton, partonic cross-sections that are calculable, and polarized fragmentation functions.
Among the hyperons, the $\Lambda$ and $\bar{\Lambda}$ hyperons are attractive probes~\cite{Xu:2002hz,Chen:2007tm} since a substantial fraction of their spin is expected to be carried by strange quarks and anti-quarks, and their hard production rate~\cite{Abelev:2006cs} is comparatively high.
Measurements of $\Lambda$ and $\bar{\Lambda}$ $D_{LL}$ can thus contribute insights into longitudinally polarized fragmentation functions and strange quark and anti-quark helicity distributions in ways that are complementary to other constraints~\cite{deFlorian:1997zj,AguilarArevalo:2010cx,deFlorian:2014yva,Nocera:2014gqa,Ethier:2017zbq,Green:2017keo,Alexandrou:2017oeh}.

In this paper we report an improved measurement of the longitudinal spin transfer $D_{LL}$ to $\Lambda$ and ${\bar \Lambda}$ hyperons in longitudinally polarized proton--proton collisions at $\sqrt{s}$ = 200\,GeV.
The data were recorded with the STAR experiment~\cite{NIM} in the year 2009 and correspond to an integrated luminosity, $\mathcal{L}$, of about 19\,pb$^{-1}$ with an average longitudinal beam polarization, $P_\mathrm{beam}$, of 57\%, measured with a relative 4.7\% accuracy~\cite{Jinnouchi:2004up,BBC:2005,Okada:2006dd,CNI:2009}.
This data sample has a figure-of-merit, $P_\mathrm{beam}^2\mathcal{L}$, approximately twelve times higher than that of our previous $D_{LL}$ measurement~\cite{Abelev:2009xg}.

The STAR subsystems used in the measurement include the Time Projection Chamber (TPC)~\cite{TPC}, which is able to track charged particles in the pseudo-rapidity range $| {\eta} |<1.3$ with full coverage in the azimuthal angle $\phi$.
Particle identification was provided via measurements of specific energy loss, $dE/dx$, due to ionization from charged particles passing through the TPC gas. 
The Barrel and Endcap Electromagnetic Calorimeters (BEMC and EEMC)~\cite{BEMC, EEMC} are lead-sampling calorimeters covering $|\eta|<1$ and $1.1<\eta < 2$, respectively, with full coverage in $\phi$.
The BEMC and EEMC were used as trigger detectors to initiate the recording of data.
Collision events were recorded if they satisfied a jet-patch trigger condition in the BEMC or EEMC, which required transverse energy deposits that exceeded thresholds of $\simeq \! 5.4\,\mathrm{GeV}$ (JP1, prescaled) or $\simeq \! 7.3\,\mathrm{GeV}$ (L2JetHigh) in a patch of calorimeter towers  spanning a range of $\Delta \eta$$\times$$\Delta \phi$ = 1 $\times$ 1 in pseudo-rapidity and azimuthal angle.
For our previous $D_{LL}$ measurements~\cite{Abelev:2009xg} the BEMC covered only $\eta > 0$ and lower trigger thresholds were used.  The EEMC was not used.

A longitudinal $\Lambda$ or $\bar{\Lambda}$ polarization component, $P_{\Lambda(\bar{\Lambda})}$,  manifests itself through a dependence of the number of observed hyperons on the angle $\theta^*$ of the decay proton or anti-proton in the hyperon rest frame in the weak decay channel $\Lambda \to p \pi^-$ or $\bar \Lambda \to \bar p \pi^+$:
\begin{equation}
\frac{dN}{d \cos{\theta}^*}=\frac{\sigma\mathcal{L}A}{2}(1+\alpha_{\Lambda(\bar{\Lambda})} P_{\Lambda(\bar{\Lambda})}
\cos{\theta}^*), 
\label{ideal}
\end{equation}
where $A$ is the detector acceptance, 
$\alpha_{\Lambda(\bar{\Lambda})}$ is the weak decay parameter, 
and $\theta^*$ is the angle
between the $\Lambda(\bar{\Lambda})$ momentum direction (i.e. longitudinal polarization) and the (anti-)proton momentum in the $\Lambda(\bar{\Lambda})$ rest frame.
The dependence of $A$ on $\theta^*$ and other observables is omitted in this notation.

The analysis methods are very similar to those of our prior $D_{LL}$ measurement~\cite{Abelev:2009xg}.
The $\Lambda$ and $\bar{\Lambda}$ candidates were identified from the topology of their dominant weak decay channels, $\Lambda \to p \pi^-$ and $\bar \Lambda \to \bar p \pi^+$, each having a branching ratio of 63.9\%~\cite{PDG}. 
TPC tracks were required to be formed by a minimum of 15 hits on the pads of the 45 TPC pad-rows.
The beam-collision vertex was reconstructed event-by-event from charged particle tracks reconstructed with the TPC.
This vertex, the primary event vertex, was required to be along the beam axis and within 60\,cm of the TPC center to ensure uniform tracking efficiency.
The data for each beam-collision event were then searched for (anti-)proton and pion tracks with curvatures of opposite sign.
The (anti-)protons and charged pions were identified by requiring that $dE/dx$ was within three standard deviations of the respective nominal values.
The tracks were then paired to form a $\Lambda(\bar{\Lambda})$ candidate and topological selections were applied to reduce combinatorial and $K_S^0$ backgrounds.
The selections included criteria for the distance of closest approach (DCA) between the paired tracks, the DCA of the reconstructed candidate track to the  primary event vertex, the DCAs of the (anti-)proton and pion tracks to the primary event vertex, the decay length of the $\Lambda(\bar{\Lambda})$ candidate, and the cosine of the angle between the $\Lambda(\bar{\Lambda})$ candidate momentum and its trajectory from the primary event vertex, $\cos(\vec{r}\cdot\vec{p})$.
These criteria were tuned in $p_T$ intervals so as to keep as much signal as possible while keeping the residual background at an acceptable level of about 10\%~\cite{thesis-cendejas}.
Table~\ref{tab:cutsV0} summarizes the track and candidate selection criteria, the number of $\Lambda(\bar{\Lambda})$ candidates used in the analysis, and the estimated residual background.
The larger number of $\bar{\Lambda}$ than $\Lambda$ in the analysis has its origins primarily in the trigger conditions and thresholds for the recorded event sample and the energy deposit in the calorimeters associated with the annihilation of anti-protons from $\bar{\Lambda}$ decay.
In minimum-bias proton--proton collisions, the $\bar{\Lambda}$ yield is below the one for $\Lambda$~\cite{Abelev:2006cs}.
\begin{table*}
\begin{center}
\setlength\extrarowheight{1pt}
\begin{tabular}{@{ }l c c c c c c}
\hline\hline
Selection criterion & 2--3\,GeV/$c$ & 3--4\,GeV/$c$ & 4--5\,GeV/$c$ & 5--8\,GeV/$c$ \tabularnewline
\hline 
\small
DCA of $p\pi^-$ ($\bar{p}\pi^+$)  & $<$ 0.7\,cm & $<$ 0.5\,cm & $<$ 0.5\,cm & $<$ 0.5\,cm \tabularnewline
DCA of $\Lambda$ ($\bar{\Lambda}$)  & $<$ 1.2\,cm & $<$ 1.2\,cm & $<$ 1.2\,cm & $<$ 1.2\,cm \tabularnewline
DCA of $p$ ($\bar p$)  & $>$ 0.2\,cm &  $-$ & $-$ & $-$ \tabularnewline
DCA of $\pi^\pm$  & $>$ 0.4\,cm & $>$ 0.4\,cm & $>$ 0.4\,cm & $>$ 0.4\,cm \tabularnewline
Decay length  & $>$ 3.0\,cm & $>$ 3.5\,cm & $>$ 4.0\,cm & $>$ 4.5\,cm \tabularnewline 
$\cos({\vec r} \cdot {\vec p})$  & $>$ 0.98  & $>$ 0.98 & $>$ 0.98  & $>$ 0.98   \tabularnewline
\hline
$\Lambda$ ($\bar{\Lambda}$) counts  & 151\,340 (243\,964) & 63\,308 (105\,564) & 23\,070 (35\,568) & 15\,642 (18\,939) \tabularnewline
$\Lambda$ ($\bar{\Lambda}$) bkgd frac. & 0.146 (0.101) & 0.114 (0.081) & 0.094 (0.072) & 0.127 (0.115) \tabularnewline
\hline\hline
\label{allcuts}
\end{tabular} 
\caption{Summary of the selection criteria used in the analysis (see text) to identify $\Lambda$ ($\bar{\Lambda}$) candidates for different intervals in $p_T$ and the number of $\Lambda$ ($\bar{\Lambda}$) candidates used in the analysis, together with the estimated fractions of residual background.}
\label{tab:cutsV0}
\end{center}
\end{table*}  
\begin{figure}
\begin{center}
\includegraphics[width=0.5\textwidth]{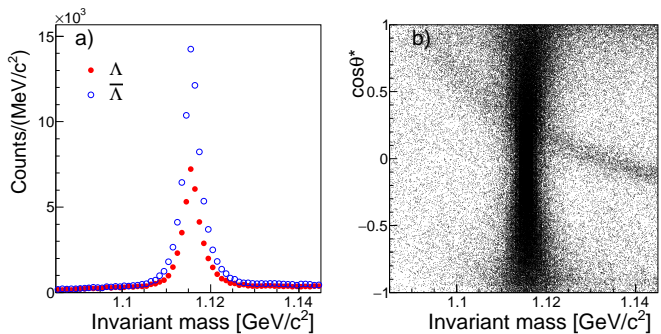}
\caption{ a) The invariant mass distribution for $\Lambda$ (red filled circles) and $\bar{\Lambda}$ (blue open circles) candidates with 3 $< p_T <$ 4\,GeV/$c$ in this analysis and b) the distribution of hyperon rest-frame angle $\cos\theta^*$ versus invariant mass $m_{\Lambda(\bar\Lambda)}$. } 
\label{mass}
\end{center}
\end{figure}

Figure~\ref{mass}a) shows the invariant mass distribution for the reconstructed $\Lambda$ (filled circles) and $\bar\Lambda$ (open circles) candidates with $|\eta_{\Lambda(\bar\Lambda)}| <$ 1.2 and 3 $<p_T<$ 4\,GeV/$c$ from the data sample obtained with the JP1 trigger condition.
Figure~\ref{mass}b) shows $\cos\theta^*$ versus invariant mass $m_{\Lambda(\bar\Lambda)}$ for these events.  Besides residual combinatorial background, the distribution shows a band originating from $K_S^0$ particle decays when a fraction of the decay pions with $p_T >$ 1.2\,GeV/$c$ is misidentified as a (anti-)proton.   
The $\Lambda$ and $\bar{\Lambda}$ candidates within the invariant mass range $1.110<m_{\Lambda(\bar\Lambda)} < 1.122\,\mathrm{GeV}/\mathrm{c}^2$ were kept for further analysis.
In addition, the $\Lambda$ and $\bar{\Lambda}$ baryons were required to be associated with a reconstructed jet that satisfied the trigger conditions.
For this purpose, a jet sample was reconstructed using the mid-point cone algorithm~\cite{Blazey:2000qt} as in several previous STAR jet analyses~\cite{Abelev:2006uq,Abelev:2007vt,Adamczyk:2012qj}.
The association required that the reconstructed $\eta$ and $\phi$ of the $\Lambda$ or $\bar{\Lambda}$ candidate were within the jet cone of radius $\Delta\mathcal{R} =\sqrt{(\Delta\eta)^2+(\Delta \phi)^2} = 0.7$.  Reconstructed jets were required to have $p_T > 5\,\mathrm{GeV/}c$.
The fraction of $\Lambda$ $(\bar{\Lambda})$ hyperons associated with a near-side trigger jet increases with increasing hyperon $p_T$ from about 43\% (55\%) for  $2 < p_T <  3\,\mathrm{GeV}/c$ to 62\% (72\%) for  $5 < p_T <  8\,\mathrm{GeV}/c$.  The larger fraction for $\bar{\Lambda}$ than for $\Lambda$ is due to the aforementioned energy deposit from annihilation of decay anti-protons in the calorimeters.

The longitudinal spin transfer $D_{LL}$ was extracted in small intervals of $\cos\theta^*$ from the ratio:
\begin{equation}
D_{LL}=\frac{1}{\alpha_{\Lambda(\bar{\Lambda})} P_\mathrm{beam} \left<\cos \theta^*\right>} \frac{N^+ -R N^-} {N^+ + R N^-},
\label{eq_dll}
\end{equation}
where $\alpha_\Lambda = 0.642\pm0.013$~\cite{PDG}, $\alpha_{\bar\Lambda} = -\alpha_\Lambda$, 
$N^+$ $(N^-)$ is the number of $\Lambda$ or $\bar{\Lambda}$ hyperons in the $\cos\theta^*$ interval when the beam is positively (negatively) polarized, $\left<\cos\theta^*\right>$ is the average value of $\cos\theta^*$ in this interval, and $R=\mathcal{L}^+/\mathcal{L}^-$ denotes the corresponding luminosity ratio for the two beam polarization states.
The detector acceptance cancels in this ratio~\cite{Abelev:2009xg}.
Eq.~\ref{eq_dll} follows from Eqs.~\ref{gener1} and~\ref{ideal} and parity conservation in the hyperon production processes.
The observed (raw) yields contain the produced hyperons as well as residual background.
In addition, both beams at RHIC are polarized.
In the analysis, the (raw) candidate yields $n^{++}$, $n^{+-}$, $n^{-+}$, and $n^{--}$ by helicity configuration of the RHIC beams were weighted with the corresponding relative luminosities to determine the single spin yields used in Eq.~\ref{eq_dll}. That is, in the analysis $N^+ = n^{++} + n^{+-}$ if the luminosities are equal and for the case that the ``first'' beam is considered polarized and the second unpolarized.  Analogously, $N^- = n^{-+} + n^{--}$ up to effects of relative luminosity.   Similar expressions hold in the case that the second beam is considered polarized and the first unpolarized.  In both cases,  forward rapidity is defined with respect to the forward-going polarized beam.
The relative luminosities were measured with Beam-Beam Counters (BBC)~\cite{BBC:2005}.
The (raw) spin transfer values were then averaged over the entire $\cos\theta^*$ range and a correction was applied for the effects from the residual backgrounds:
\begin{equation}
D_{LL}=\frac{D_{LL}^\mathrm{raw}-rD_{LL}^\mathrm{bg} }{1-r},
\label{eq:DLL}
\end{equation}
where the fraction of residual background,  $r$, within the accepted mass interval $1.110 < m_{\Lambda(\bar\Lambda)} < 1.122\,\mathrm{GeV}/\mathrm{c}^2$, and the spin transfer for the residual background, $D_{LL}^\mathrm{bg}$, were estimated using side-bands $1.094 < m_{\Lambda(\bar\Lambda)} < 1.103\,\mathrm{GeV}/\mathrm{c}^2$ and $1.129 < m_{\Lambda(\bar\Lambda)} < 1.138\,\mathrm{GeV}/\mathrm{c}^2$ on either side of the $\Lambda$ or $\bar{\Lambda}$ mass peak.
Simulation shows that less accurate results are obtained if this correction procedure is applied in each $\cos\theta^*$ interval and the resulting $D_{LL}(\cos\theta^*)$ values are then combined.
$D_{LL}^\mathrm{bg}$ was found to be consistent with zero to within its statistical uncertainties.
The statistical uncertainties in the background-corrected $D_{LL}$ values were calculated according to:
\begin{equation}
\delta D_{LL} = \frac{
\sqrt{ \left( \delta D^\mathrm{raw}_{LL} \right)^2 + \left(r \delta D^\mathrm{bg}_{LL}\right)^2}
}{1-r},
\label{eq:DLL-uncertainty}
\end{equation}
which thus contains $r$ and the statistical uncertainty in background $D_{LL}^\mathrm{bg}$.
The uncertainty in $r$ is accounted for in a contribution to the systematic uncertainty.

\begin{figure}
\begin{center}
\includegraphics[width=0.5\textwidth]{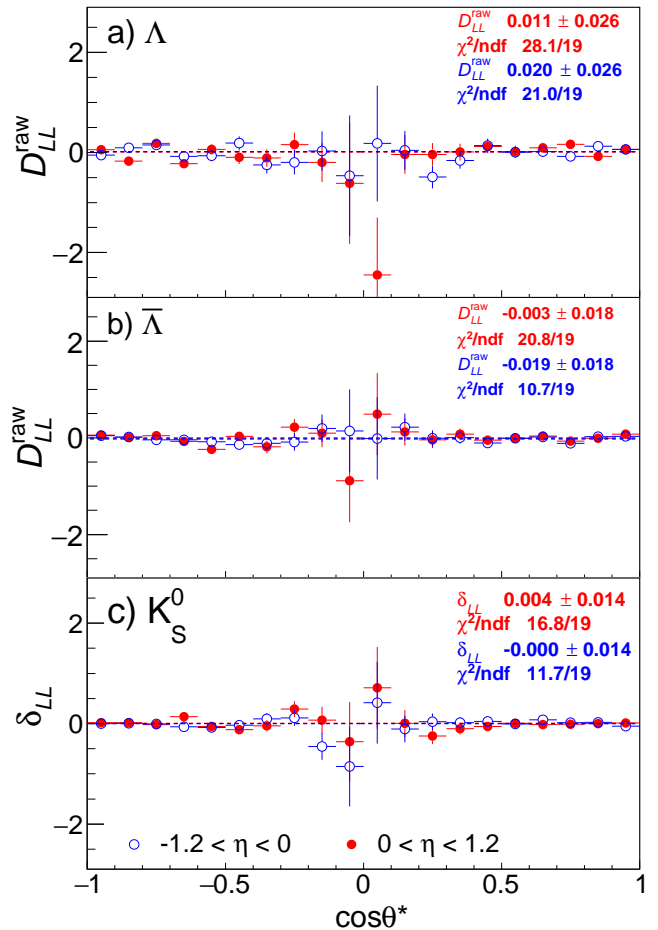}
\caption{The spin transfer $D^\mathrm{raw}_{LL}$ versus $\cos\theta^*$ for a) $\Lambda$ and b) $\bar{\Lambda}$ hyperons, and c) the spin asymmetry $\delta_{LL}$ for the control sample of $K_\mathrm{S}^0$ mesons versus $\cos\theta^*$ for $3 < p_T < 4\,\mathrm{GeV/}c$ from JP1 triggered data. The red filled circles show the results for positive pseudorapidity $\eta$ 
and the blue open circles show the results for negative $\eta$. Only statistical uncertainties are shown.}
\label{cosin_fit}
\end{center}
\end{figure}

Figure~\ref{cosin_fit}a) shows $D^\mathrm{raw}_{LL}$ obtained from Eq.~\ref{eq_dll} versus $\cos\theta^*$ for $\Lambda$ baryons with 3 $<p_T<$ 4\,GeV/$c$ for intervals of positive and negative pseudorapidity with respect to the momentum of the incident polarized proton.  Only statistical uncertainties are shown and the data satisfied the JP1 trigger condition.  Figure~\ref{cosin_fit}b) shows the corresponding $\bar{\Lambda}$ results.
The $\Lambda$ and $\bar{\Lambda}$ results are constant with $\cos\theta^*$, as expected and as confirmed by the fit quality of the averages.
A null-measurement was performed by analyzing the spin transfer to the spinless $K_\mathrm{S}^0$ meson, $\delta_{LL}$, through the $K_S^0 \to \pi^+\pi^-$ decay channel.  This decay channel has a topology similar to the $\Lambda \to p \pi^-$ and $\bar\Lambda \to \bar p \pi^+$ channels.
The values for $\delta_{LL}$ were determined with an artificial weak decay parameter $\alpha = 1$ using otherwise identical methods as for the hyperon spin transfer measurements.
The results are shown in Fig.~\ref{cosin_fit}c) and are consistent with zero as expected.
No significant asymmetries $A_L$, defined as the cross-section asymmetry for positive and negative beam helicity in single polarized proton--proton scattering, were observed either, as expected at $\sqrt{s} = 200\,\mathrm{GeV}$.
The asymmetries $A_{LL}$, defined as the cross-section asymmetry for aligned and opposed beam helicity configurations in double polarized proton--proton scattering, do not necessarily vanish.
While no statistically significant values were observed for the $\Lambda$ and $\bar{\Lambda}$ hyperons, an average value of $A_{LL} = 0.006 \pm 0.002$ was observed for $K^0_\mathrm{S}$ mesons associated with jets for $p_T > 1\,\mathrm{GeV/}c$.
%

%
\begin{figure*}
\begin{center}
  \includegraphics[width=0.75\textwidth]{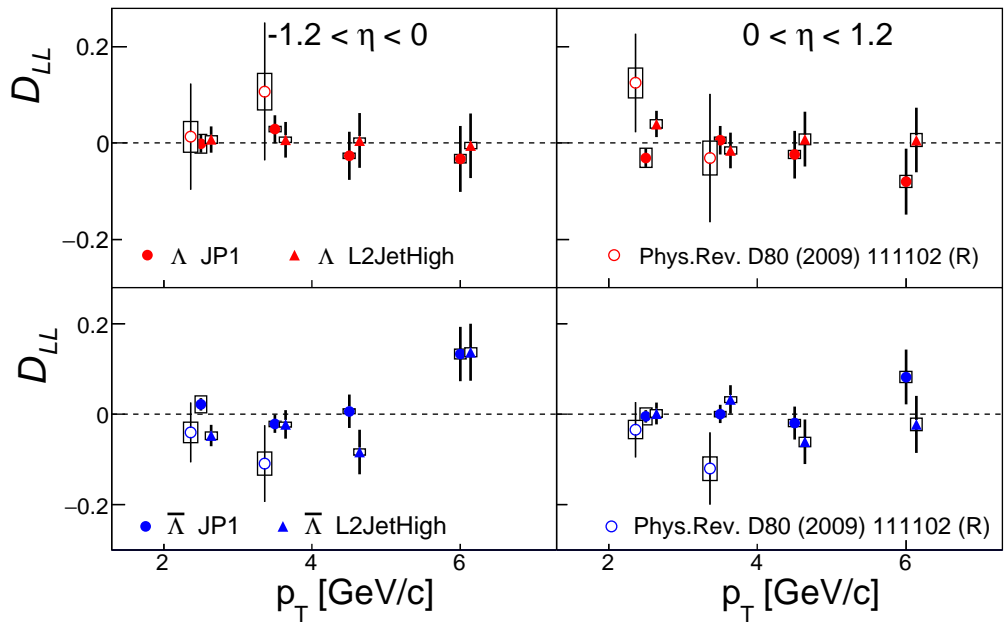}
\caption{Comparison of spin transfer $D_{LL}$ for  positive and negative $\eta$ versus $p_T$ for differently triggered data samples in the present analysis, together with previously published results in the region of kinematic overlap.  The vertical bars and boxes indicate the sizes of the statistical and systematic uncertainties, respectively. The results obtained with the L2JetHigh trigger have been offset to slightly larger $p_T$ values for clarity. The previously published results have been offset to slightly smaller $p_T$ values.}
\label{DLL09_2trg_etap}
\end{center}
\end{figure*}

Figure~\ref{DLL09_2trg_etap} shows a comparison of the results of $D_{LL}$ to the $\Lambda$ (top) and $\bar{\Lambda}$ (bottom) for negative (left) and positive (right) hyperon pseudorapidities obtained from the JP1 and L2JetHigh triggered data samples in comparison with previously published data~\cite{Abelev:2009xg} in the region of kinematic overlap.  The error bars show the size of the statistical uncertainties, while the boxes indicate the size of the total systematic uncertainty.  The central values along the $x$-axis have been shifted slightly to higher $p_T$ values for the L2JetHigh data for visual clarity, while the previously published results have been offset to slightly smaller values.
The present data are seen to surpass the prior results in precision and kinematic range.

\begin{table*}
\begin{center}
\setlength\extrarowheight{1pt}
\begin{tabular}{c@{\ \ \ }c@{\ \ }cc@{\ \ }c@{\ \ }c}
\hline\hline
$p_T$                      & \multicolumn{2}{c}{$\Lambda$} & &  \multicolumn{2}{c}{$\bar{\Lambda}$} \\ \cline{2-3} \cline{5-6}
$[\mathrm{GeV}/c]$ & $\eta < 0$ & $\eta > 0$ & & $\eta < 0$ & $\eta >0$ \\
\hline 
\small
2.4 & $\ \ \, 0.002 \pm 0.015 \pm 0.016$  &  $-0.008 \pm 0.015 \pm 0.016$ & & $\ \ \, 0.005 \pm 0.011 \pm 0.016$  &  $      -0.003 \pm 0.011 \pm 0.016$ \\
3.4 & $\ \ \, 0.021 \pm 0.022 \pm 0.005$  & $-0.002 \pm 0.022 \pm 0.007$ & & $      -0.022 \pm 0.016 \pm 0.006$  &  $\ \ \, 0.010 \pm 0.016 \pm 0.006$ \\
4.4 & $      -0.013 \pm 0.036 \pm 0.005$  & $-0.010 \pm 0.036 \pm 0.009$ & & $      -0.025 \pm 0.028 \pm 0.006$  & $      -0.034 \pm 0.028 \pm 0.008$ \\
5.9 & $      -0.019 \pm 0.048 \pm 0.008$  & $-0.036 \pm 0.048 \pm 0.013$ & & $\ \ \, 0.135 \pm 0.043 \pm 0.010$  & $\ \ \, 0.032 \pm 0.043 \pm 0.013$ \\
\hline\hline
\end{tabular} 
\caption{Measured $D_{LL}$ values for $\Lambda$ and $\bar{\Lambda}$ hyperons at different $p_T$ and $\eta$ with statistical and systematic uncertainties.}
\label{tab:results}
\end{center}
\end{table*} 

The size of the total systematic uncertainties ranges from 0.006 to 0.017, varying with $p_T$.
The improvement in overall size compared to our previous previous $D_{LL}$ measurement~\cite{Abelev:2009xg} is due mostly to the refined treatment of background (c.f. Eqs.~\ref{eq:DLL} and~\ref{eq:DLL-uncertainty}) made practicable with the larger data sample.
The size of the systematic uncertainty was estimated by considering contributions from uncertainties in the decay parameter, the beam polarization, residual transverse beam polarization components, the relative luminosities, as well as contributions from uncertainties in the fraction of residual background, uncertainties caused by event overlap (pileup) in the detector, and uncertainties introduced by the trigger conditions~\cite{thesis-cendejas}.
Among these, the dominant sources of the systematic uncertainty are from pileup and from trigger bias.
These causes of systematic uncertainty are uncorrelated and their effects act primarily as offsets to the data.
The effects of pileup were studied with the data by considering variations of the reconstructed spin-sorted hyperon yields per beam-collision event for different collision rates.  The reconstructed hyperon yield per collision event is expected to be constant in the absence of pileup.
Constant and linear extrapolations to small collision rates, where pileup vanishes, were then used to estimate the contribution from any existing pileup effects in the data to the systematic uncertainty.  The resulting uncertainty contribution is found to be largest for small values of $p_T$.
The trigger conditions can affect the composition of the recorded data sample in several ways.  For example, it could change the relative fractions of the underlying hard scattering processes.
The trigger conditions can also distort the sampling for different momentum fractions in the fragmentation.
In addition, the trigger conditions could affect the contributions to the $\Lambda$ or $\bar{\Lambda}$ yields from the decays of heavier hyperons.
Each of these effects was studied with Monte Carlo simulated events that were generated with PYTHIA 6.4.28~\cite{pythia}, using the Perugia 2012 tune~\cite{Skands:2010ak} further adapted to the  conditions at RHIC~\cite{thesis-chang} and an increased K-factor~\cite{Abelev:2006cs}.
These events were then passed through the STAR detector response package based on GEANT 3~\cite{Geant}.
The effects from differences caused by the trigger conditions were then evaluated using the model expectations for $D_{LL}$ from Ref.~\cite{Xu:2005ru}.  Their size was found to increase with $p_T$ and forms the dominant contribution to the systematic uncertainty at large $p_T$.

\begin{figure}
\begin{center}
\includegraphics[width=0.49\textwidth]{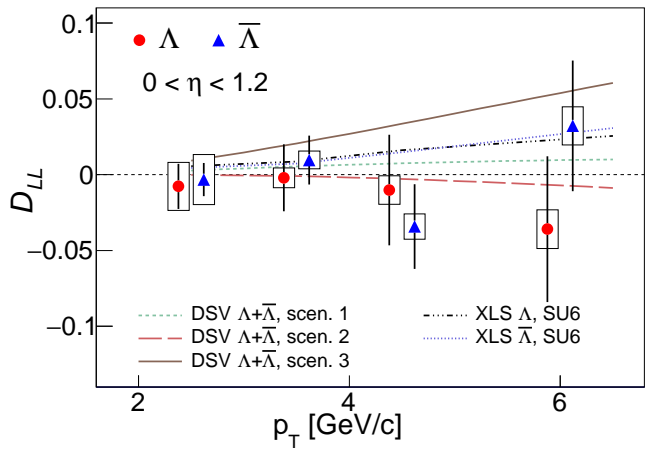}
\caption{Comparison of the measured spin transfer $D_{LL}$ with theory predictions for positive $\eta$ versus $p_T$. The vertical bars and boxes indicate the sizes of the statistical and systematic uncertainties, respectively. The $\bar{\Lambda}$ results have been offset to slightly larger $p_T$ values for clarity.}
\label{DLL_09etap}
\end{center}
\end{figure}

The results for the different trigger conditions in Fig.~\ref{DLL09_2trg_etap} are point-by-point consistent, as demonstrated by $\chi^2 = 17$ for 16 degrees of freedom.  The results from the JP1 and L2JetHigh trigger conditions were thus combined, using statistical weights.
Figure~\ref{DLL_09etap} shows the combined data on the spin transfer $D_{LL}$ to the $\Lambda$ and to the $\bar{\Lambda}$ as functions of $p_T$ for positive $\eta$.
The data provide no evidence for a difference between  $\Lambda$ and $\bar{\Lambda}$ $D_{LL}$, as indicated by $\chi^2 = 1.5$ for 4 degrees of freedom.
The curves show the theory expectations from Ref.~\cite{deFlorian:1998ba,PrivCom}, which considers $D_{LL}$ for $\Lambda$ and $\bar\Lambda$ combined, and from Ref.~\cite{Xu:2005ru}, which considers $D_{LL}$ separately for $\Lambda$ and for $\bar\Lambda$.
The theory expectations are seen to increase in size with increasing $p_T$.  They increase also with increasing $\eta$.
The differences between the curves from Ref.~\cite{deFlorian:1998ba,PrivCom} arise primarily from assumptions for the polarized fragmentation functions, which are thus far only poorly constrained.
The data do not provide evidence for a non-vanishing spin transfer signal; their comparison with zero
yields $\chi^2 = 3$ for 8 degrees of freedom.
However, the data tend to be below a theory expectation based on the extreme assumption that the quark polarized fragmentation functions are flavor-independent, commonly referred to as the DSV scenario 3 expectation~\cite{deFlorian:1998ba,PrivCom}.
The overall probability for DSV scenario 3 to yield a dataset with all central values anywhere below the expectation is less than 1\% if eight data points are considered and about 6\% for four data points.  This corresponds to what is seen from the data if the $\Lambda$ and $\bar{\Lambda}$ points are considered separately and if the $\Lambda$ and $\bar{\Lambda}$ data are combined for each $p_T$ value.
Table~\ref{tab:results} contains the numerical values of the data points in Fig.~\ref{DLL_09etap} as well as the corresponding data for negative $\eta$.
STAR has recently made the first measurement of the transverse spin transfer, $D_{TT}$, for $\Lambda$ and $\bar{\Lambda}$ hyperons produced in transversely polarized proton--proton collisions~\cite{Adam:2018wce}.  $D_{TT}$ is sensitive to the quark transversity distributions.
In addition, STAR is expanding its acceptance by means of an ongoing upgrade to the inner sectors of the TPC and has proposed an instrument upgrade that would enable a program of measurements, including $D_{LL}$ and other $\Lambda$ and $\bar{\Lambda}$ measurements, at very forward rapidities~\cite{Aschenauer:2016our}.

In summary, we report an improved measurement of the longitudinal spin transfer, $D_{LL}$, to $\Lambda$ hyperons and $\bar{\Lambda}$ anti-hyperons in longitudinally polarized proton--proton collisions at $\sqrt{s}$ = 200\,GeV.  The data correspond to an integrated luminosity of 19\,pb$^{-1}$ with an average beam polarization of 57\% and were obtained with the STAR experiment in the year 2009.  The $\Lambda$ and $\bar{\Lambda}$ data cover  $|\eta|<$ 1.2 and $p_T$ up to 6\,GeV/$c$.  The longitudinal spin transfer is found to be $D_{LL}$ = -0.036 $\pm$ 0.048\,(stat) $\pm$ 0.013\,(sys) for $\Lambda$ hyperons and $D_{LL}$ = 0.032 $\pm$ 0.043\,(stat) $\pm$ 0.013\,(sys) for $\bar{\Lambda}$ anti-hyperons produced with $\left<\eta\right>$ = 0.5 and $\left<p_T\right>$ = 5.9\,GeV/$c$, where the corresponding theory expectations reach their largest sizes.  While the data do not provide conclusive evidence for a spin transfer signal, the data tend to be below a theory expectation, DSV scenario 3~\cite{deFlorian:1998ba,PrivCom}, based on the extreme assumption that the quark polarized fragmentation functions are flavor-independent.

We thank the RHIC Operations Group and RCF at BNL, the NERSC Center at LBNL, and the Open Science Grid consortium for providing resources and support.  This work was supported in part by the Office of Nuclear Physics within the U.S. DOE Office of Science, the U.S. National Science Foundation, the Ministry of Education and Science of the Russian Federation, National Natural Science Foundation of China, Chinese Academy of Science, the Ministry of Science and Technology of China and the Chinese Ministry of Education, the National Research Foundation of Korea, Czech Science Foundation and Ministry of Education, Youth and Sports of the Czech Republic, Department of Atomic Energy and Department of Science and Technology of the Government of India, the National Science Centre of Poland, the Ministry  of Science, Education and Sports of the Republic of Croatia, RosAtom of Russia and German Bundesministerium f\"ur Bildung, Wissenschaft, Forschung and Technologie (BMBF) and the Helmholtz Association.

\end{document}